\documentstyle[12pt]{article}
\advance\hoffset by -7mm
\makeatletter
\@addtoreset{equation}{section}
\makeatother

\renewcommand{\title}[1]{\null\vspace{25mm}
\renewcommand{\thefootnote}{\fnsymbol{footnote}}
\noindent{\Large{\bf #1}}\vspace{10mm}

\noindent {\large By }}

\renewcommand{\abstract}[1]{\vspace{10mm}

\noindent{\small{\em Abstract.} #1}\vspace{2mm}

}

\begin{document}
\def\be{\begin{equation}}
\def\ee{\end{equation}}
\def\bea{\begin{eqnarray}}
\def\eea{\end{eqnarray}}
\def\bean{\begin{eqnarray*}}
\def\eean{\end{eqnarray*}}
\def\ba{\begin{array}} \def\ea{\end{array}}
\newcommand{\f}{\frac}
\newcommand {\equ}[1] {(\ref{#1})}

\def\6{\partial} \def\a{\alpha} \def\b{\beta}
\def\g{\gamma} \def\d{\delta} \def\ve{\varepsilon}
\def\e{\epsilon}
\def\z{\zeta} \def\h{\eta} \def\th{\theta}
\def\vt{\vartheta} \def\k{\kappa} \def\l{\lambda}
\def\m{\mu} \def\n{\nu} \def\x{\xi} \def\p{\pi}
\def\r{\rho} \def\s{\sigma} \def\t{\tau}
\def\Ph{\phi} \def\ph{\varphi} \def\ps{\psi}
\def\o{\omega} \def\G{\Gamma} \def\D{\Delta}
\def\Th{\Theta} \def\L{\Lambda} \def\S{\Sigma}
\def\PH{\Phi} \def\Ps{\Psi} \def\O{\Omega}
\def\sm{\small} \def\la{\large} \def\La{\Large}
\def\LA{\LARGE} \def\hu{\huge} \def\Hu{\Huge}
\def\ti{\tilde} \def\wti{\widetilde}
\def\non{\nonumber\\}
\def\={\!\!\!&=&\!\!\!}
\def\+{\!\!\!&&\!\!\!+~}
\def\-{\!\!\!&&\!\!\!-~}
\def\id{\!\!\!&\equiv&\!\!\!}
\renewcommand{\AA}{{\cal A}}
\newcommand{\BB}{{\cal B}}
\newcommand{\CC}{{\cal C}}
\newcommand{\DD}{{\cal D}}
\newcommand{\EE}{{\cal E}}
\newcommand{\FF}{{\cal F}}
\newcommand{\GG}{{\cal G}}
\newcommand{\HH}{{\cal H}}
\newcommand{\II}{{\cal I}}
\newcommand{\JJ}{{\cal J}}
\newcommand{\KK}{{\cal K}}
\newcommand{\LL}{{\cal L}}
\newcommand{\MM}{{\cal M}}
\newcommand{\NN}{{\cal N}}
\newcommand{\OO}{{\cal O}}
\newcommand{\PP}{{\cal P}}
\newcommand{\QQ}{{\cal Q}}
\newcommand{\RR}{{\cal R}}
\newcommand{\SS}{{\cal S}}
\newcommand{\TT}{{\cal T}}
\newcommand{\UU}{{\cal U}}
\newcommand{\VV}{{\cal V}}
\newcommand{\WW}{{\cal W}}
\newcommand{\XX}{{\cal X}}
\newcommand{\YY}{{\cal Y}}
\newcommand{\ZZ}{{\cal Z}}
\newcommand{\R}{R^{j}_{\alpha_{0}}}
\newcommand{\Z}{Z^{\alpha_{0}}_{\alpha_{1}}}
\newcommand{\journal}[4]{{\em #1~}#2\,(19#3)\,#4;}
\newcommand{\aihp}{\journal {Ann. Inst. Henri Poincar\'e}}
\newcommand{\hpa}{\journal {Helv. Phys. Acta}}
\newcommand{\sjpn}{\journal {Sov. J. Part. Nucl.}}
\newcommand{\ijmp}{\journal {Int. J. Mod. Phys.}}
\newcommand{\physu}{\journal {Physica (Utrecht)}}
\newcommand{\pr}{\journal {Phys. Rev.}}
\newcommand{\jetpl}{\journal {JETP Lett.}}
\newcommand{\prl}{\journal {Phys. Rev. Lett.}}
\newcommand{\jmp}{\journal {J. Math. Phys.}}
\newcommand{\rmp}{\journal {Rev. Mod. Phys.}}
\newcommand{\cmp}{\journal {Comm. Math. Phys.}}
\newcommand{\cqg}{\journal {Class. Quantum Grav.}}
\newcommand{\zp}{\journal {Z. Phys.}}
\newcommand{\np}{\journal {Nucl. Phys.}}
\newcommand{\pl}{\journal {Phys. Lett.}}
\newcommand{\mpl}{\journal {Mod. Phys. Lett.}}
\newcommand{\prep}{\journal {Phys. Reports}}
\newcommand{\prepsec}{\journal {Phys. Reports (Review Section
                                of Phys. Letters)}}
\newcommand{\ptp}{\journal {Progr. Theor. Phys.}}
\newcommand{\nc}{\journal {Nuovo Cim.}}
\newcommand{\app}{\journal {Acta Phys. Pol.}}
\newcommand{\apj}{\journal {Astrophys. Jour.}}
\newcommand{\apjl}{\journal {Astrophys. Jour. Lett.}}
\newcommand{\annp}{\journal {Ann. Phys. (N.Y.)}}
\newcommand{\anp}{\journal {Ann. of Phys.}}
\newcommand{\Nature}{{\em Nature}}
\newcommand{\PRD}{{\em Phys. Rev. D}}
\newcommand{\MNRAS}{{\em M. N. R. A. S.}}

\begin{titlepage}
\rightline{UMTG-187}
\vskip 2cm
\centerline{\Large \bf Koszul-Tate Cohomology} 
\centerline{\Large \bf For an Sp(2)-Covariant Quantization}
\centerline{\Large \bf of Gauge Theories With Linearly Dependent Generators}
\vskip 2cm
\centerline{\sc Liviu T\u{a}taru}
\centerline{Department of Theoretical Physics}
\centerline{University of Cluj}
\centerline{Mihail Kog\u{a}lniceanu no.1}
\centerline{Cluj-Napoca, Romania, 3400}
\centerline{ltatar@hera.ubbcluj.ro}
\centerline{and}
\centerline{\sc Radu T\u{a}tar}  
\centerline{University of Miami}
\centerline{Department of Physics}
\centerline{Coral Gables, Florida, USA, 33124}
\centerline{tatar@phyvax.ir.miami.edu}
\vskip 2cm
\centerline{\sc Abstract}
\vskip 0.2in

The anti-BRST transformation,
in its Sp(2)-symmetric version, for the
general case of any stage-reducible gauge
theories is implemented in the usual BV \cite{bv1,bv2} approach. This task is
accomplished not by duplicating the gauge symmetries but rather by
duplicating all fields and antifields of the theory and by imposing the
acyclicity of the Koszul-Tate differential. In this way the
Sp(2)-covariant quantization can be realised in the standard BV approach and its
equivalence with BLT quantization \cite{blt1,blt2} 
can be proven by a special gauge fixing
procedure.
\vskip 0.4in
\end{titlepage} 

\newpage
\section{Introduction}
\setcounter{equation}{0}
Without any doubt the most popular and powerful method for the covariant
quantization of the gauge systems is the Batalin-Vilkovisky(BV) method
\cite {bv1,bv2}, which uses the nice mathematical structure of Poisson
brackets (in fact, antibrackets), canonical transformations, etc. i.e.,
the attactive features of the hamiltonian approach in the
Lagrangian approach, keeping the advantages of a covariant formalism.
However, in the standard BV method only the minimal sector of the
theory occurs quite natural from the acyclicity of the
Koszul-Tate differential \cite{h}, and the non-minimal sector, which is
crucial for the applications, is difficult to be understood and fixed.

The main purpose of Batalin, Lavrov and Tyutin (BLT) Sp(2)-quantization
\cite{blt1,blt2,blt3} of the gauge theories is to offer a proper understanding
of the {\em non-minimal sector}, which becomes a natural part of the
minimal sector. Nevertheless, the BLT method, in spite of the fact that it
is very similar to BV method, has  different structure and a different
field content. First of all the sympectic structure of the theory is lost
since the usual antibrackets are replaced by
the extended antibrackets and the master equarion is replaced by
two generating equations \cite{blt1,blt2,blt3}. The field structure of the
theory is quite asymmetric since a field $\phi^A$ is associated with
{\em two} antifields $\Phi^*_{Aa},a=1,2$ and a bar-field
$\bar{\Phi}_A$, so the symmetry field $\leftrightarrow$ antifield
is lost in this case. Furthermore, the gauge fixing  process is
quite tricky and it seems to be difficult to the connection, if any,
between BV method and BLT method.

In this paper we shall show that it is possible to reformulate the
anti-BRST, Sp(2) formalism in the usual BV framework
just by {\em duplicating} all the fields and antifields of the
theory and by using
homological perturbation theory (HPT) \cite{fhst}. The main ingredient of the
 HPT is the construction of {\em the Koszul-Tate (KT) differential}
 $\delta_K$ and its  {\  acyclicity}. This differential and its properties
 have been studied details in \cite{fhst} (see also \cite{fh}) in order to prove
 the existence and the uniqueness for the solution
 of the master equation. The {\em acyclicity} of KT differential
 determines, in fact, the spectrum of all antifields and therefore of
 all  fields, whether we work within a theory with a
 symmetric field-antifield structure. In our approach we
 shall work only with pairs of field-antifield and we shall
 reduce the BLT quantization for the reducible systems to the
 usual one.

 What is basic in our construction is the remark that if we duplicate
 the antifields, they form a redundant basis of vectors
 and in order to identify the algebra of polynomials in
 the fields and antifields with the algebra
 of multivectors it is necessary to set
 $ \Phi^*_{A1}=\Phi^*_{A2}$ in the former algebra, where $\Phi_{Aa}, a=1,2$
 are all antifields from the theory. The most natural way to
 accomplish this task is to suppose
 that {\em the full BRST differential} does this job
  ( and not only KT differential). This assumption is quite strong and it
 implies that all fields (incdluding the ghosts ) must occur in pairs and the
 action must depend only on the
 {\em sum} of these pairs i.e.
 \be
 S=S(\Phi^{A1}+\Phi^{A2}, \cdots)
\label{ac}
 \ee
 where the index A is common index for all fields and ghosts which
 occur in our theory. But the action \ref{ac} has an additional gauge symmetry
 $\Phi^{Aa} \rightarrow  \Phi^{Aa}+ (-1)^a\epsilon^A ,a=1,2$,
 which must be taken into account, if we want to
 quantize properly the theory. Thus durring the
 $Sp(2)$ quantization, in our version, we have to duplicate the fields
 and antifields and to consider the new gauge symmetry which we have
 just mentioned. This has already been done by one of us \cite{lt1}
 in the irreducible case. Here we intend to extend our
construction for the general case
 of the reducible systems.

 \section{The spectrum of antifields}
\setcounter{equation}{0}
Let us consider a classical set of fields
$\Phi^j (j=1,\cdots ,n=n_++n_-),$  where $n_+(n_-)$ is the number of
Boson (Fermion) fields. The classical
action $S_0(\Phi)$ is supposed to be invariant under the gauge transformation

\bea
\Phi^j \rightarrow \Phi^{\prime j} & = & \Phi^j+\delta\Phi^j=\Phi^j+R^j_{\alpha_0}
(\Phi)\xi^{\alpha_0} \nonumber\\
G_j R^j_{\alpha_0}(\Phi) & = & S_{0,j}(\Phi)R^j_{\alpha_0}(\Phi)=0,
\hspace{1cm} \alpha_0=1,\dots, m_0, m_0=m_{0+}+m_{0-},
\eea
where $\xi^{\alpha_0}$ are the parameters of the
gauge transformations, with the Grassman parity
$\epsilon(\xi^{\alpha_0})=\epsilon_{\alpha_0}$;
\hspace{0.2cm} $\epsilon(R^{j}_{\alpha_{0}})=\epsilon_j+\epsilon_{\alpha_0}$
and $\epsilon(\Phi^{i})=\epsilon_i$ , $G_j=S_{0,j}$ the comma here means
the derivative of $S_0$ with respect of $\Phi^j$.

For the case of $\mbox{rank}(R^j_{\alpha_0})= (r_{0+},r_{0-})$ with
$r_{0\pm}=m_{0\pm}$  when the generators $R^j_{\alpha_0}$ are all
independent, the structure and the quantization of the theory have been
already discussed in \cite{lt1}.

If $r_{0\pm}<m_{0\pm}$ the generators $R^j_{\alpha_0}$ are linearly
dependent and the theory is {\em reducible}. In this case
the matrix $R^j_{\alpha_0} $ has on mass shell $S_{0,j}=0$
a number of $m_1=m_{1+}+m_{1-}$ zero-eigenvalue eigenvectors $Z^{\alpha_0}
_{\alpha_1}$ and the numbers $\epsilon_{\alpha_1}=0,1$ such that:
\be
R^i_{\alpha_0}Z^{\alpha_0}_{\alpha_1}=S_{0,j}L^{ij}_{\alpha_1},
\hspace{1cm}\alpha_1=1,\cdots,m_1;
\ee
with $\epsilon(Z^{\alpha_0}_{\alpha_1})=\epsilon_{\alpha_0}+
\epsilon_{\alpha_1}$ and the matrices $L^{ij}_{\alpha_1}$ can
be chosen to have the properties
\be
L_{\alpha_1}^{ij}=-(-1)^{\epsilon_i\epsilon_j}L_{\alpha_1}^{ji}.
\ee
If the $rank Z^{\alpha_0}_{\alpha_1}=(r_{1+}, r_{1-}) $ with
$r_{1\pm}<m_{1\pm}=m_{0\pm}-r_{0\pm}$ the set $\{Z^{\alpha_0}_{\alpha_1}\}$
is linearly dependent and there is  the set  of zero-eigenvalue
eigenvectors $Z^{\alpha_1}_{\alpha_2}$ and the numbers
$\epsilon_{\alpha_2}=0,1$ such that
\be
Z^{\alpha_0}_{\alpha_1}Z^{\alpha_1}_{\alpha_2}=S_{0,j}L^{\alpha_0 j}_{\alpha_2},
\hspace{1cm}\alpha_2=1,\cdots,m_2;
\ee
with $\epsilon(Z^{\alpha_1}_{\alpha_2})=\epsilon_{\alpha_1}+
\epsilon_{\alpha_2}$
and the matrices $L^{\alpha_0j}_{\alpha_2}$
can be chosen to have the properties
\be
L_{\alpha_2}^{\alpha_0 j}=-(-1)^{\epsilon_{\alpha_0}\epsilon_j}
L_{\alpha_2}^{j \alpha_0}.
\ee

In the general case the set $\{Z^{\alpha_1}_{\alpha_2}\}$ could
be redundant and we have to continue the process. Thus we have a
sequence of reducibility equations for the sets $\{Z_{\alpha_s}^{\alpha_{s-1}}
\}, \hspace{0.2cm} (s=1,\cdots, L) $ of the form:
\be
Z_{\alpha_{s-1}}^{\alpha_{s-2}}Z_{\alpha_{s}}^{\alpha_{s-1}}=
S_{0,j} L_{\alpha_s}^{\alpha_{s-2} j},\hspace{0.1cm} \alpha_s=1,
\cdots,m_s=m_{s+}+m_{s-}, \hspace{0.2cm} s=1,\cdots, L.
\ee

In order to find the spectrum of fields and  ghosts of our theory
we shall use a natural way, which differs from the one chosen by
Gr\'{e}goire and Henneaux \cite{gh1} since we will start with the
spectrum of the antifields and we define the spectrum of fields
just by a simple correspondence  antifield $\rightarrow $ field.
On the other hand, the spectrum of the antifields is uniquely determined ,
by the demand that the Koszul-Tate differential $\delta_K$
be acycle. This fact was emphasised by Fisch and Henneaux \cite{fh}
in their atempt to clarify the algebraic structure of the
antifield-antibracket formalism for
reducible gauge theories in the usual Batalin-Vilkovisky quantization.
They showed that the acyclicity of $\delta_K$ forces the antifield
spectrum to be just the correct minimal set described
by Batalin and Vilkovisky \cite{bv1,bv2}(see also \cite{h}).

We shall use the general scheme developed by Fisch and
Henneaux but we are going to introduce a new ingredient in the
theory, which will allow us to duplicate all the fields and
antifields and to  obtain, in one way, the correct set of fields and
antifields described by Batalin, Lavrov and Tyutin in their
Sp(2)-covariant quantization of gauge theories
with liniar dependent generators \cite{blt2,blt3}. However, in our approach
not only the spectrum of all antifields is correctly obtained but
also the form of the quantized action is determined.

At the irreducible level,we start with a doublet of fields $\Phi_{ia}$,
where $a=1,2$.The action will depend only on the sum of these fields.
The Koszul-Tate differential acting on the antifields associated to these
fields is given by
\be
\delta_{K}\Phi^{*}_{ja}=-S_{0,ja}=-\frac{\delta S_{0}}{\delta\Phi^{ja}}
\ee
where $S_{0} = S_{0}(\Phi^{j1} + \Phi_{j2})$. By subtracting the relations given
by the previous equation, we obtain $\delta_{K}(\Phi^{*}_{j1}-\Phi^{*}_{j2})=0$,
i.e. we have a cycle. But all the cycles have to be trivial, so we introduce a
new field $\bar{\Phi}_{i}$ whose differential is equal with the previous cycle
which becomes now a boundary. We thus have
\be
\delta_{K}\bar{\Phi}_{i}=\Phi^{*}_{j1}-\Phi^{*}_{j2}
\ee
Hitherto, we did not use the symmetries of the theory. We will proceed now to
consider them. The Noether theorem tells us that it exist $R^{i}_{\a}$, such
that $R^{i}_{\a} S_{0,ia} = 0$, which can be written in another form, as
$\delta_{K}(R^{j}_{\a_{0}}\Phi^{*}_{ja}) = 0$, which introduce another cycle  
which
has to be killed. We kill this cycle by introducing antifiels of higher
antighost, which will have a supplementary indice, $C^{*}_{\a_{0}b\vert a}$,
with
\be
R^{j}_{\alpha_{0}}\Phi^{*}_{ja}=\delta_{K}C^{*}_{\alpha_{0}b\vert a}
\ee
We have again new cycles,given by a specific combination of the antifields
just introduced,given by
\be
\delta_{K}(C^{*}_{\alpha_{0}1\vert a} - C^{*}_{\alpha_{0}2\vert a}) = 0
\ee
and
\be
\delta_{K}(C^{*}_{\alpha_{0}2\vert 1} - C^{*}_{\alpha_{0}1\vert 2} -
R^{j}_{\alpha_{0}}\bar{\Phi}_{j}) = 0
\ee
which again have to be killed. To realize this, one has to introduce new
antifields, given by the relations
\be
C^{*}_{\alpha_{0}1\vert a}-C^{*}_{\alpha_{0}2\vert a}=\delta_{K}\bar{C}_
{\alpha_{0}\vert a}
\ee
and
\be
C^{*}_{\alpha_{0}2\vert 1}-C^{*}_{\alpha_{0}1\vert 2}-
R^{j}_{\alpha_{0}}\bar{\Phi}_{j}=\delta_{K}B^{*}_{\alpha_{0}a}
\ee
where the fields $\bar{C}_{\alpha_{0}\vert a}$ and $\bar{B}_{\alpha_{0a}}^{*}$
 are introduced in order to kill the nontrivial cycles.

With this we have solved the irreducible theory, i. e. we do not have any
nontrivial cycle and we introduced the entire spectrum of antifields. To
all the antifields we associate a field, therefore we have besides the
previous antifields the fields  $\bar{\bar{\Phi}}\Leftrightarrow \bar{\Phi}$,
$\bar{\bar{C}}^{\alpha_{0}\vert a}\Leftrightarrow \bar{C}_{\alpha_{0}\vert a}$,
$C^{\alpha_{0}b\vert a}\Leftrightarrow C^{*}_{\alpha_{0}b\vert a}$,
$B^{\alpha_{0}a}\Leftrightarrow B^{*}_{\alpha_{0}a}$, where the arrows denote
the correspondence between the respective fields and antifields. The total
action, which includes all the fields and the antifields is given by
\begin{eqnarray}
S = \Phi^{*}_{ja}R^{j}_{\alpha_{0}}(C^{\alpha_{0}1\vert a}
+ C^{\alpha_{0}2\vert a})\\ \nonumber
+ (C^{*}_{\alpha_{0}2\vert 1} - C^{*}_{\alpha_{0}1\vert 2} -
R^{j}_{\alpha_{0}}\bar{\Phi}_{j})R^{j}_{\alpha}(B^{\alpha_{1}} + B^{\alpha_{2}})
+ \bar{\bar{\Phi}}^{A}(\Phi^{*}_{A1}-\Phi^{*}_{A2})\
\end{eqnarray}
where $\Phi^{A}$ is a collective notation for all the fields.

In this paper we assume that the tensorial character of the ghost 
$C^{\alpha_{0}1}$ and the antighost $C^{\alpha_{0}2}$ is the same, the
last one belonging to the non-minimal sector in the standard version of BV.
This fact is true only for certain types of gauge fixings, so we work
by stating this restriction. One example where our approach does not work is the
bosonic string in conformal gauge where the ghost and the antighost have 
different transformations under diffeomorphisms.

We go now to the reducible theory. We call reducible a theory where there
exists a relation between the R functions, i. e. there exist functions Z , 
such that
$R^{j}_{\alpha_{0}}Z^{\alpha_{0}}_{\alpha_{1}} = 0$. In this case
the appearance of new cycles which have to be killed by Koszul-Tate
differential becomes obvious. They are given by:

1. $Z^{\alpha_{0}}_{\alpha_{1}}C^{*}_{\alpha_{0}b\vert a}$ which is to be
brought to a boundary by introducing new antifields with the relation
\be
\frac{1}{2}Z^{\alpha_{0}}_{\alpha_{1}}(C^{*}_{\alpha_{0}b\vert a} +
C^{*}_{\alpha_{0}a\vert b}) = \delta_{K}C^{*}_{\alpha_{1}c\vert ab}
\ee
where we imposed the symmetry in a and b indices, a condition required by
the Sp(2) symmetry.

2. $C^{*}_{\alpha_{1}1\vert ab} - C^{*}_{\alpha_{1}2\vert ab}$, which is to be
killed by a new bar field introduced with
\be
C^{*}_{\alpha_{1}1\vert ab}-C^{*}_{\alpha_{1}2\vert ab}=\delta_{K}
(\bar{C}_{\alpha_{1}\vert ab})
\ee

3. $C^{*}_{\alpha_{1}2\vert 1b}-C^{*}_{\alpha_{1}1\vert 2b}+
\bar{C}_{\alpha_{0}\vert b} Z^{\alpha_{0}}_{\alpha_{1}}-
\frac{1}{2}B^{*}_{\alpha_{0}b}Z^{\alpha_{0}}_{\alpha_{1}}$, which is killed
by introducing a new antifield as:
\be
C^{*}_{\alpha_{1}2\vert 1b}-C^{*}_{\alpha_{1}1\vert 2b}+
\bar{C}_{\alpha_{0}\vert b} Z^{\alpha_{0}}_{\alpha_{1}}-
\frac{1}{2}B^{*}_{\alpha_{0}b}Z^{\alpha_{0}}_{\alpha_{1}}=\delta_{K}
B^{*}_{\alpha_{1}c\vert b}
\ee

4. $B^{*}_{\alpha_{1}1\vert b}-B^{*}_{\alpha_{1}2\vert b}$ which is killed by a
new bar field given by
\be
B^{*}_{\alpha_{1}1\vert b}-B^{*}_{\alpha_{1}2\vert b}=\delta_{K}
\bar{B}_{\alpha_{1}\vert b}
\ee
These are all the antifields which have to be added for a first stage reducible
theory. Besides, we have to introduce all the fields which correspond to these
antifields with the same correspondence as we used for the irreducible theory.

Using the results for irreducible and first-stage reducible theories, we can
now give the formula for general theories, i. e. for the s-stage reducible
theories. For these theories, there exist $Z^{\alpha_{i-1}}_{\alpha_{i}}$,
with i from 0 to s, such that $Z^{\alpha_{i-1}}_{\alpha_{i}}
Z^{\alpha_{i}}_{\alpha_{i+1}}=0$, which are the reducibility conditions.

By induction, we have obtained the following results for the action of the
Koszul-Tate differential on the antifields:
\be
\delta_{K}C^{*}_{\alpha_{s}b\vert a_{1}\cdots a_{s+1}}=\frac{1}{s+1}
\SS[C^{*}_{\alpha_{s-1}a_{1}\vert a_{2}\cdots a_{s+1}} Z^{\alpha_{s-1}}_
{\alpha_{s}}]
\ee
and
\begin{eqnarray}
\delta_{K}B^{*}_{\alpha_{s}b\vert a_{1}\cdots a_{s}}=-\epsilon^{ac}
C^{*}_{\alpha_{s-1}a\vert a_{1}\cdots a_{s}}- \nonumber \\
-\frac{1}{s+1}Z^{\alpha_{s-1}}_
{\alpha_{s}}\SS [B^{*}_{\alpha_{s-1}a_{1}\vert a_{2}\cdots a_{s}}]-
\bar{C}_{\alpha_{s-1}\vert a_{1}\cdots a_{s}}Z^{\alpha_{s-1}}_
{\alpha_{s}}\
\end{eqnarray}
where $\epsilon^{ab}=0$ if $a\not= b$, 1 if $a=1,b=2$ and -1 for $a=2, b=2$
and the symbol $\SS$ denotes the symetrisation over the indices which appear
after the vertical line in the formulas for C and B because of the Sp(2)
symmetry. Moreover we have,
\be
\delta_{K}\bar{C}_{\alpha_{s}\vert a_{1}\cdots a_{s}}=
C^{*}_{\alpha_{s}1\vert a_{1}\cdots a_{s}}
-C^{*}_{\alpha_{s}2\vert a_{1}\cdots a_{s}}
\ee
To all the previously introduced antifields we assign fields using the same
conventions as for irreducible and first stage reducible theories.

The minimal action, i. e. the action constructed only with the minimal sector
is given by:
\begin{eqnarray}
S & = & S_{0}+\sum_{s=0}\frac{1}{s+1}
\S [C^{*}_{\alpha_{s-1}a_{1}\vert a_{2}\cdots a_{s+1}} Z^{\alpha_{s-1}}_
{\alpha_{s}}](C^{\alpha_{s}1\vert a_{1}\cdots a_{s}}
+C^{\alpha_{s}2\vert a_{1}\cdots a_{s}})] \nonumber \\
&- & (\epsilon^{ac}
C^{*}_{\alpha_{s-1}a\vert a_{1}\cdots a_{s}}+\frac{1}{s+1}Z^{\alpha_{s-1}}_
{\alpha_{s}}S[B^{*}_{\alpha_{s-1}a_{1}\vert a_{2}\cdots a_{s}}] +
\bar{C}_{\alpha_{s-1}\vert a_{1}\cdots a_{s}}Z^{\alpha_{s-1}}_
{\alpha_{s}})(B^{\alpha_{s}1\vert a_{1}\cdots a_{s}}
+B^{\alpha_{s}2\vert a_{1}\cdots a_{s}} \nonumber \\
& + & {\bar{\bar{\Phi}}}^{A}(\Phi^{*}_{A2}-\Phi^{*}_{A1}) \
\label{boundary}
\end{eqnarray}

The master equation in our case, as was discussed in \cite{ltt},
 is written as
\be
\label{master}
(S_{T}~,~S_{T})=0
\ee
where
\be
S_{T}=\bar{\bar{\Phi}}^{A}(\Phi^{*}_{A2}-\Phi^{*}_{A1})+
S(\Phi^{A},\Phi^{*}_{Aa},\bar{\Phi}_{A})
\ee
where again $\Phi^{A}$ is the symbol for all the fields i.e.
$\Phi^{A} = \Phi^{A1} + \Phi^{A2}$.

\section{The Sp(2)-BRST transformations for a first stage reducible system}
\setcounter{equation}{0}
Consider now a first-stage reducible system, which is caracterized by
the functions 
$R^{j}_{\alpha_{0}},Z_{\alpha_{1}}^{\alpha_{0}}$, with the reducibility
relations $R^{j}_{\alpha_{0}} Z_{\alpha_{1}}^{\alpha_{0}}\equiv 0$.
This corresponds to an off-shell reducible theory and in the sequel we will
discuss only this type of reducible theories, which have the property
of being linear in antifields. For an on-shell reducible theory, the solution
becomes at least quadratic in antifields which makes it more difficult to solve
(in the case of Freedman - Townsend model the solution was obtained by
Barnich et al. in \cite{bcg}). 

Moreover,we have the commutativity relation between R's:
\be
\label{adi}
[R_{\alpha_{0}},R_{\beta_{0}}]=C^{\gamma_{0}}_{\alpha_{0}\beta_{0}}
R^{j}_{\gamma_{0}}
\ee
where $C^{\gamma_{0}}_{\alpha_{0}\beta_{0}}$ are constants giving the algebra
of transformations.

The action, which is a solution of the master equation (\ref{master})
with the boundary conditions (\ref{boundary}), for the one-reducible theories
with a closed algebra exists as a linear functional in the antifields
$\PH^*_{Aa}$ and $\bar{\PH}_A$ and has the general form
\be
\label{general}
S_T=S_0(\PH)+\bar{\bar{\PH}}^A (\PH^*_{A1}-\PH^*_{A2})+\PH^*_{Aa}X^{Aa}+
\bar{\PH}_A Y^A
\ee
where $X^{Aa}$ and $Y^A$ are funtions of the fields $\PH^A=\PH^{A1}+\PH^{A2}$.
If one defines the following transformations of the fields
\be
s^a \PH^A  =X^{Aa}
\label{transf}
\ee
then the master equation (\ref{master}) yields
\bea
s^a S_0&=& 0 \non
\{s^a~,~s^b\}\PH^A &=& 0 \non
Y^A &=& \f1{2}\e_{ab}s^a s^b \PH^A  \non
s^a Y^A &=& 0
\eea

Starting from these data,we can follow very closely the results of Spiridonov
\cite{s}, who determined the transformation $s^a \PH^A$  of fields.
Adopting his results to our notations, and {\em using everywhere $\Phi^{A}$
instead of $\Phi^{A1}+\Phi^{A2}$ }, the results
are
\bea
s^{a}\Phi^{j} & = & R^{j}_{\alpha_{0}}C^{\alpha_{0}\vert a}, \non
s^{a}C^{\alpha_{0}\vert b} & = & \epsilon^{ab}B^{\alpha_{0}}-
\frac{1}{2}C^{\alpha_{0}}_{\beta_{0}\gamma_{0}}C^{\beta_{0}\vert a}
C^{\gamma_{0}\vert b}+Z^{\alpha_{0}}_{\alpha_{1}}C^{\alpha_{1}\vert ab}, 
\label{transf2}
\eea

\bea
s^{a}B^{\alpha_{0}} &=& Z^{\alpha_{0}}_{\alpha_{1}}B^{\alpha_{1}\vert a}
+\non
&+&\frac{1}{2}C^{\alpha_{0}}_{\beta_{0}\gamma_{0}}(B^{\beta_{0}}C^{\gamma_{0}
\vert a}
+Z^{\beta_{0}}_{\alpha_{1}}C^{\alpha_{1}\vert ab}\epsilon^{bc}
C^{\gamma_{0}\vert c})\nonumber \\
&-&\frac{1}{12}(C^{\alpha_{0}}_{\beta_{0}\gamma_{0}}C^{\beta_{0}}_{\sigma_{0}\rho_{0}}
+2 C^{\alpha_{0}}_{\sigma_{0}\rho_{0},j}R^{j}_{\gamma_{0}})C^{\sigma_{0}\vert a}
C^{\rho_{0}\vert b}\epsilon^{bc} C^{\gamma_{0}\vert c}.\
\label{transf3}
\eea

These relations were constructed only with the variable
$C^{\alpha_{0}}_{\sigma_{0}\rho_{0},j}$ and are for
the fields which appear in the case of an irreducible theory. For the rest
of fields, one needs new variables,which are combinations of R, Z and C's.

The corresponding transformations are
\begin{eqnarray}
& s^{a} & C^{\alpha_{1}\vert bc}=-\epsilon^{ab}B^{\alpha_{1}\vert c}-
\epsilon^{ac}B^{\alpha_{1}\vert b}\nonumber \\
& + & A^{\alpha_{1}}_{\alpha_{0}\beta_{1}}C^{\alpha_{0}\vert a} C^{\beta_{1}
\vert bc}
-\frac{1}{2}F^{\alpha_{1}}_{\alpha_{0}\beta_{0}\gamma_{0}}C^{\alpha_{0}\vert a}
C^{\beta_{0}\vert b}C^{\gamma_{0}\vert c}\
\label{transf5}
\end{eqnarray}
and
\begin{eqnarray}
\label{transf6}
s^{a}B^{\alpha_{1}\vert b} & = & A^{\alpha_{1}}_{\alpha_{0}\beta_{1}}C^{\alpha_{0}a}
B^{\beta_{1}\vert b}-\frac{1}{2}F^{\alpha_{1}}_{\alpha_{0}\beta_{0}\gamma_{0}}
B^{\alpha_{0}}C^{\beta_{0}\vert a}C^{\gamma\vert b}\nonumber \\
 & + & \frac{1}{2}A^{\alpha_{1}}_{\alpha_{0}\beta_{1}}Z^{\alpha_{0}}_{\alpha_{1}}
C^{\alpha_{1}\vert ac}C^{\gamma_{1}\vert bd}\epsilon^{cd}
-\frac{1}{4}F^{\alpha_{1}}_{\alpha_{0}\beta_{0}\gamma_{0}}Z^{\alpha_{0}}_{\beta_{1}}
(3C^{\beta_{0}\vert a}C^{\beta_{1}\vert bc}\epsilon^{cd} C^{\gamma_{0}\vert d}
\nonumber \\ 
& + & C^{\beta_{0}\vert b}C^{\beta_{1}\vert ac} \epsilon^{cd} C^{\gamma_{0}
\vert d}) \
\end{eqnarray}

It is worthwhile to note that the BRST differential s defined by
\be
sF = (F , S_{T})
\ee
acts on $\Phi^{A} = \Phi^{A1} + \Phi^{A2}$ as $s\Phi^{A} = s^{1}\Phi^{A}
+ s^{2} \Phi^{A}$. Therefore, if one considers only the cohomology of s
on the algebra generated by fields $\Phi^{A}$ and their derivatives then we 
can define three cohomology groups H(s), H($s_{1}$) and H($s_{2}$). 
Remembering that any free differential algebra (A, D) can be decomposed into
a contractive subalgebra generated by elements of the form (x, Dx) and
a minimal algebra M (with the condition $DM \subset M\times M$), the cohomology
group of (A, D) is given by H(M, D) because the contractive part does not
contribute. Using this observation and the previous transformation of the
fields under $s^{1}, s^{2}$, it results that the non-minimal of the
original BV scheme $C^{\alpha_{0}\vert 2}, B^{\alpha_{0}}, C^{\alpha_{1}
\vert 12}, B^{\alpha_{1} \vert 2}$ do not give any contribution to 
H($s_{1}$). It also results that $C^{\alpha_{0}\vert 1}, B^{\alpha_{0}},
 C^{\alpha_{1}
\vert 12}, B^{\alpha_{1} \vert 1}$ do not give any contribution to
the cohomology H($s_{2}$). This proves the compatibility of our approach with
the triviality of the non-minimal sector in the usual BV approach ( see
for example \cite{bhh}).
 
Here, the A's and F's which appear in the equations are given by the folowing
relations:
\begin{equation}
\label{doina}
Z^{\alpha_{0}}_{\alpha_{1},j}R^{j}_{\beta_{0}}+C^{\alpha_{0}}_{\beta_{0}\gamma_{0}}
Z^{\gamma_{0}}_{\alpha_{1}}+Z^{\alpha_{0}}_{\beta_{1}}A^{\beta_{1}}_{\beta_{0}
\alpha_{1}}=0
\end{equation}
or
\begin{equation}
C^{\alpha_{0}}_{\beta_{0}\gamma_{0}} Z^{\beta_{0}}_{\alpha_{1}}
Z^{\gamma_{0}}_{\beta_{1}}+A^{\gamma_{1}}_{\beta_{0}\beta_{1}}
Z^{\beta_{0}}_{\alpha_{1}}Z^{\alpha_{0}}_{\gamma_{1}}=0
\label{doina2}
\end{equation}
and
\begin{eqnarray}
A^{\alpha_{1}}_{\alpha_{0}\beta_{1}}A^{\beta_{1}}_{\beta_{0}\gamma_{1}}-
A^{\alpha_{1}}_{\beta_{0}\beta_{1}}A^{\beta_{1}}_{\alpha_{0}\gamma_{1}}+
A^{\alpha_{1}}_{\alpha_{0}\gamma_{1},j}R^{j}_{\beta_{0}} \nonumber \\
-A^{\alpha_{1}}_{\beta_{0}\gamma_{1},j}R^{j}_{\alpha_{0}}+
A^{\alpha_{1}}_{\gamma_{0}\gamma_{1}} C^{\gamma_{0}}_{\alpha_{0}\beta_{0}}+
3F^{\alpha_{1}}_{\alpha_{0}\beta_{0}\gamma_{0}}Z^{\gamma_{0}}_{\gamma_{1}}=0 \
\label{doina3}
\end{eqnarray}
In the original paper of Spiridonov,the fields $B^{a}$ appeared as auxiliary
fields which were used as Lagrange multipliers for the gauge conditions.In
our approach,the B fields are associated to the antifields introduced with
the acciclicity of the Koszul-Tate differential.In \cite{s} the F fields were
introduced starting from the equation (\ref{adi}) which expresses the
closure of the gauge algebra which was assumed in our approach.If we aply the
Jacoby identity in the case of field-dependent structure constants for
first-stage reducible theory we can introduce the F fields by writing the
Jacobi identity in the following form:
\begin{equation}
\label{adi1}
C^{\alpha_{0}}_{\beta_{0}\gamma_{0}}C^{\gamma_{0}}_{\rho_{0}\sigma_{0}}+
R^{j}_{\beta_{0}}C^{\alpha_{0}}_{\rho_{0}\sigma_{0},j}+
Z^{\alpha_{0}}_{\alpha_{1}}F^{\alpha_{1}}_{\beta_{0}\rho_{0}\sigma_{0}}+
(\beta_{0}\rho_{0}\sigma_{0}-cycle)
\end{equation}
where $\beta_{0}\rho_{0}\sigma_{0}$-cycle means the circular permutation
of the indices $\beta_{0}, \rho_{0}, \sigma_{0}$.

The A's coefficients are obtained by taking (\ref{adi}), multiplying it by
$Z^{\alpha_{0}}_{\alpha_{1}}$ and using:
\be
R^{j}_{\alpha_{0},k}Z^{\alpha_{0}}_{\alpha_{1}}
=-R^{j}_{\alpha_{0}}Z^{\alpha_{0}}_{\alpha_{1},k}
\ee
The result of is then introduced into (\ref{doina}) which
gives the new structure coefficients denoted by A.

In the section 3 we have obtained the action of the BRST operator (which is
the Koszul-Tate operator for the case of antifields) on antifields
and which is a generalisation of the result of \cite{lt1} for a theory
of any order reducibility. In section 4 we have made the parallel between our
approach and Spiridonov approach for the action of the BRST operator on the
fields in the case of one-reducible theory.

\section{Gauge fixing in our approach}
\setcounter{equation}{0}
In the newest version of the BRST-BV quantization \cite{bt,bms,bm}
the path integral is proposed to be
\be
Z=\int d\G d\l e^{\f{i}{\hbar}(S+X)}
\label{z}
\ee
where $\G$ denotes the collection of all fields and antifields of the theory
and $X(\G, \l)$ is a hypergauge fixing action, depending on a new set of
variables $ \l^A$ with $\e(\l^A)=\e_A$, which can be considered as
Lagrange multipliers, when $X$ depends linear in them. In order to assure the
gauge independence of the path integral $Z$ (\ref{z}), $X$ must
satisfy the master equation
\be
\label{master2}
(X~,~X)=0.
\ee
 In the BV quatization the hypergauge action has the form
\be
\label{standard}
X=\l^A(\PH^*_A -\f{\6\Psi}{\6\PH^A})
\ee
which verifies the master equation (\ref{master2}) for any $\l^A$.

In the extended Sp(2) theory we have three types of fields
$\PH^{A1}~,~\PH^{A2}~,~\bar{\bar{\PH}}^{A}$ and three types of
antifields $\PH_{A1}^*~,~\PH_{A2}^*~,~\bar{\PH}_{A}$. We want to eliminate
the fields $\PH^{A2}$ in the path integral. Therefore we could chose
$X$ to be
\be
X=\m_A \PH^{A2} +\p^A (\PH^*_{A1}-\f{\6 \Psi}{\6 \Ph^{A1}})+\l^A(\bar{\PH}_A -
\f{\6\Psi}{\6\bar{\bar{\PH}}^A})
\ee
where $\m_A~,~\p^A$ and $\l^A$ are the Lagrange multipiers and
$\Psi=\Psi(\PH~,~\bar{\bar{\PH}})$ is the fermion gauge fixing function.

Due to this particular form of the hypergauge action the master
equation (\ref{master2}) for $X$ yields
\be
\f{\6^2\Psi}{\6\PH^{A1}\6\bar{\bar{\PH}}^B}=\f{\6^2\Psi}
{\6\PH^{B1}\6\bar{\bar{\PH}}^A}.
\ee
These equations have a solution of the form
\be
\Psi=\f{\6 F}{\6 \Ph^{A1}}\bar{\bar{\PH}}^A.
\ee
Now if we integrate out in $Z$ the variables $\m_A$ and $\l^A$ we get
eventually
\bea
S_{eff}=S(\PH ,\PH^*_1 ,\PH^*_2 ,\bar{\PH} ) - \bar{\bar{\PH}}^A
\PH^*_{A2} -\p^A \PH^*_{A1} + \non
\p^A \f{\6^2 F}{\6 \PH^A \6 \PH ^B} \bar{\bar{\PH}}^B +(\bar{\PH}_A -
\f{\6 F}{\6 \PH^A} )\l^A.
\eea
At this point we believe that it is worthwhile to remark that this form of the
quantum action coincides with the one proposed by Batalin, Lavrov and Tyutin
\cite{blt1} and by Batalin and Marnelius \cite{bm} with the identifications
$$
\p^A\rightarrow \p^{A1}~~;~~\bar{\bar{\PH}}\rightarrow \p^{A2}.
$$
but the roles played by the auxiliary fields $\p^{Aa} $ in BLT
approach are quite different from the corresponding ones in our approach.

\section{Example-topological field theories}
\setcounter{equation}{0}
As an illustration of the general theory we shall consider the
Sp(2)-quantization of the topological Yang-Mills theory. This
problem has been solved by Gomis and Roca \cite{gr} in the usual BV approach
(see however the remark at the end of our disscution) and by Perry and Teo
\cite{pt} in the usual BRST quantization (see also Baulieu \cite{bau}).

In the sequel we shall use the notations employed by Perry and Teo \cite{pt}.
In these notations the fields $\PH^j$ are the $\GG$-valued Yang-Mills 1-form
potential $A$, the ghosts $c^{\a_0 \mid a}$ are the $\GG$-valued ghosts
0-forms $c$ and $\bar{c}$ and $\GG$-valued ghost 1-forms $\psi$ and
$\bar{\psi}$, the ghosts for ghosts  $c^{\a_1\mid ab}$ are  the $\GG$-valued
ghosts 0-forms $\PH~\bar{\PH}$ and the auxiliary $\GG$-valued ghosts 0-fors
 $\l$. The additional fields $B^{\a_0}$ are the Lagrange multipliers
$b$ and $k$ and the fields $B^{\a_1\mid a}$ are the odd scalar
$\GG$-valued fields $\eta$ with the ghost number one and its
corresponding anti-ghost $\bar{\eta}$. We have found convenient to change
a little this notation and our notation is given in the Table.
\newpage 

\begin{center}
Table

\vspace{1cm}
\begin{tabular} {*{2}{|c}|c|}
\hline
General & TQFT & TQFT \\
theory & our & Perry Teo \\
       & notations & notations \\
\hline
$\PH^j$ & A     &  A \\
\hline
$c^{\a_0\mid a}$ & $c_a~\psi_a$ &$ c~\bar{c}$~\\
& & $ \psi~\bar{\psi}$\\
\hline
$c^{\a_1\mid ab}$ & $c_{ab}$ &$ \PH~\bar{\PH}~\l$\\
\hline
$B^{\a_0}$ & b~,k & b~,k\\
\hline
$B^{\a_1\mid a}$&$ \eta_a$ &$ \eta~, \bar{\eta}$\\
\hline
\end{tabular}
\end{center}

The generators of the gauge transformations $R^j_{\a_0}$ and the coefficients
$C^{\a_0}_{\b_0\g_0}$, $A^{\a_1}_{\a_0 \b_1}$ and $F^{\a_1}_{\a_0\b_0\g_0}$
can be obtained from the infinitesimal gauge transformation
\be
\label{gt}
\d A =\e_1 -D\e
\ee
with $\e_1$ an infinitesimal $\GG$-valued 1-form and $\e$ an infinitesimal
$\GG$-valued 0-form and the definitions (\ref{adi}),(\ref{doina}),
(\ref{doina2}) and (\ref{doina3}).
The gauge generators are off-shell linearly dependent: if we take
$\epsilon_1=D\tau$ and $\e=\tau$, we find that the gauge
transformation generated by $\tau$ is 0. Thus we  have a first-stage reducible
theory, for which we can apply our procedure developed in the previous
sections. We will skip the details and give only the forms of the
transformations (\ref{transf2}), (\ref{transf3}),
(\ref{transf5}) and(\ref{transf6}) for this model:
\bea
s_a A &=&\psi_a -D c_a \\
s_a c_b &=&\e_{ab} b -[c_a ,c_b ] +c_{ab}\\
s_a \psi_b &=&\e_{ab} k -[c_a , \psi_b ]-Dc_{ab}\\
s_a c_{bc}&=&-\e_{ab}\eta_c -\e_{ac}\eta_b +[c_a , c_{bc} ]\\
s_a b &=& - \eta_a + [b ,c_a ] +
\frac{1}{2}[c_{ab} ,c_c]\e^{bc} -\f1{12}[c_b ,[c_c ,c_a ]]
\e^{bc}\\
s_a k &=&  D\eta_a +[k ,c_a ] - [c_{ab} , \psi_{c}] \e^{bc} \\
s_a \eta_b &=& (-2) [c_a ,\eta_b ] +\frac{1}{2}[c_{ac} , c_{bd} ]\e^{cd}
\eea
 The quantum action  is given by the equation (\ref{general}) adapted for
the topological field theory.

Now we believe that it is noteworthy to connect our results with the
ones obtained by Gomis and Roca \cite{gr}. They have considered an usual,
i. e. not an extended Sp(2), BV action $S$ as a solution of  
the master equation and
they tried to find out an additional term  $\bar{S}$, which was the generator
of the anti-BRST transformations. These two generators can be obtained
also in the Sp(2)-BRST quantization in the situation when the solution
of the master equation involves only terms {\em linear} in the
antifields. In this case the solution of the master equation for a general
system, irreducible or reducible, in the Sp(2)-BRST quantization has the form
(again we use the notation $\PH^A =\PH^{A1} +\PH^{A2}$ ):
\be
S_T =S_0 (\PH) +\PH^*_{Aa} X^{Aa} +\bar{\PH}_A Y^A +\bar{\bar{\PH}}^A (
\PH^*_{A1}-\PH^*_{A2}).
\label{total}
\ee
In this case, it is easy to verify that the action
\be
S=S_0(\PH) +\PH^*_{A1} X^{A1}
\ee
as well as the anti-BRST  generator
\be
\bar{S}= \PH^*_A X^{A2}
\ee
obtained from \equ{total} by identifying $\PH^*_{A1}=\PH^*_{A2}=\PH^*_A$, 
fulfill the equations
\be
(S~,~S)=(S~,~\bar{S})=(\bar{S}~,~\bar{S})=0.
\ee
On the other hand in the Sp(2)-BRST quantization the ghost numbers of
$\PH^*_{A1}$ and $\PH^*_{A2}$ are different:
\be
gh(\PH^*_{A1})=-1 -gh (\PH^A)~~,~~gh(\PH^*_{A2})=1 -gh (\PH^A).
\ee
Thus the identification $\PH^*_{A1}=\PH^*_{A2}=\PH^*_A$ changes the ghost
number of $\bar{S}$ which becomes
\be
gh(\bar{S})=+2
\ee
and consequentely we can not identify $\bar{S}$ with the part of the
quantum action but rather with an anti-BRST generator.

Finally we want to remark that our resultats do not coincide neither
with the ones found by Gomis and Roca  \cite{gr} nor with the one found
out by Perry and Teo \cite{pt} but they are
{\em related by a canonical transformation} in the antifield-antibracket
formalism. In the standard theory all the anti-ghosts i.e. 
$c^{\alpha_{0}\vert 2} = \bar{c}^{\alpha_{0}}$,
$c^{\alpha_{1}\vert 22} = \bar{c}^{\alpha_{1}}$,
$c^{\alpha_{1}\vert 12} = \bar{c}'^{\alpha_{1}}$,
belong to the contractive part of the BRST free differential algebra.
We show now that the action (6.10), (6.11) can be transformed into the
standard one via a canonical transformation defined by:
\be
\phi'^{A} = \frac{\partial F_{2}(\phi, \phi'^{*})}{\partial \phi'^{*}_{A}};
\phi^{*}_{A} = \frac{\partial F_{2}(\phi, \phi'^{*})}{\partial \phi^{A}};
\ee
where $\epsilon (F_{2}) = 1$ and gh($F_{2}$) = -1. This canonical 
transformation preserves the anti-bracket structure (see \cite{tnp} ). Thus
in order to define a canonical transformation one should give the
function $F_{2}(\phi, \phi'^{*})$. 

For the actions (6.10), (6.11) the canonical transformation is generated by:

\bea
F_{2} = {\phi'}_{j}^{*} \phi^{j} + {C'}^{\alpha_{0}\vert 1 *} C_{\alpha_{1}\vert 1}
+
{C'}^{\alpha_{0}~2~*} C_{\alpha_{0}\vert 2} +
{B'}_{\alpha_{0}}^{*} s C^{\alpha_{0}\vert 2} + 
{C'}^{\alpha_{1}\vert 11 *} C_{\alpha_{1}\vert 11} + \non
{C'}^{\alpha_{1}\vert 22 *} C_{\alpha_{1}\vert 22}
{C'}^{\alpha_{1}12*} C_{\alpha_{1}\vert 22} + {B'}_{\alpha_{1}\vert 1}^{*}~
s~C^{\alpha_{1}\vert~12} + {B'}_{\alpha_{1}\vert 2}^{*} s C^{\alpha_{1}\vert22}
\eea

The form of the action (6.10) takes a simple form after the canonical
transformation:
\bea
S = S_{0} + \phi_{j}^{*} R_{\alpha_{0}}^{j} C^{\alpha_{0}\vert~1} + 
C_{\alpha_{0}\vert~1}^{*} ( Z_{\alpha_{1}}^{\alpha_{0}} C^{\alpha_{1}
~\vert~11} + \frac{1}{2}
C^{\alpha_{0}}_{\beta_{0}\gamma_{0}} C^{\gamma_{0}\vert~1} 
C^{\beta_{0}\vert~1}) + \non
+C_{\alpha_{0}\vert~2}^{*} B_{\alpha_{0}} + 
C_{\alpha_{1}\vert~12} B^{\alpha_{1}\vert~1} +
C_{\alpha_{1}\vert~22} B^{\alpha_{1}\vert~2} + \non
C_{\alpha_{1}\vert~11}^{*} ( A_{\beta_{1}~\alpha_{0}}^{\alpha_{1}} 
C^{\alpha_{0}
~\vert~1} + 
F^{\alpha_{1}}_{\alpha_{0}\beta_{0}\gamma_{0}} C^{\gamma_{0}\vert~1} 
C^{\beta_{0}\vert~1} C^{\alpha_{0}\vert 1})
\eea
 which is the standard form of the action in the BV scheme.

On the other hand, the action $\bar{S}$ takes a form 
which is much more complicated.
Since it does not play a fundamental role, we will not give it here.
The only thing that we want to emphasize is that it coincides with 
the action $\bar{S}$ obtained in \cite{gr} by using different methods.

\section{Conclusions}
\setcounter{equation}{0}
We started from the results of \cite{lt1} for irreducible gauge theories and
we applied the same ideas for the reducible theories.Our results were
similar with those of Batalin-Lavrov-Tyutin for the spectrum of
antifields and we obtained the spectrum of fields of Spiridonov directly
from cohomology methods and not by introducing auxiliary fields as Lagrangian
multiplier.Our method dealt with the cycles by introducing auxiliary
variables such that these are to be killed.

We used as an example for our approach the topological field theories and
we connected our results with the ones derived in \cite{gr},\cite{pt}.

\section{Aknowledgements}
\setcounter{equation}{0}
Liviu T\u{a}taru would like to thank to W.Kummer from Technische
Universitat Wien for an extended collaboration.

\end{document}